\documentclass[preprint,12pt]{elsarticle}

\usepackage{amsmath} \allowdisplaybreaks
\usepackage{amssymb}

\usepackage{soul,color}
\usepackage{amsfonts}
\usepackage{amstext}

\usepackage{subcaption}

\usepackage{mathtools}

\usepackage{algorithm}

 \usepackage{hyperref}

\usepackage{listings}
\usepackage{xcolor}

\usepackage{eurosym}
\usepackage{url}
\usepackage{enumitem}
\usepackage{verbatim}
\usepackage{bm}
\usepackage{pgfplots}
\usepackage[eulergreek]{sansmath}

\usetikzlibrary{positioning}
\usepgfplotslibrary{groupplots}
\pgfplotsset{compat=1.15}
\usetikzlibrary{calc}
\usepackage{relsize}
\definecolor{rosso}{RGB}{220,57,18}
\definecolor{giallo}{RGB}{255,153,0}
\definecolor{blu}{RGB}{102,140,217}
\definecolor{verde}{RGB}{16,150,24}
\definecolor{viola}{RGB}{153,0,153}
\definecolor{darkblue}{RGB}{73, 52, 235}
\definecolor{lightgreen}{RGB}{128, 235, 52}
\definecolor{cuspurple}{RGB}{128, 52, 235}
\definecolor{lighpink}{RGB}{235, 52, 204}
\definecolor{cusbrown}{RGB}{128, 110, 73}

\newcommand{\inblue}[1]{\textcolor{blue}{#1}}
\newcommand{\inred}[1]{\textcolor{red}{#1}}
\newcommand{\ingreen}[1]{\textcolor{green}{#1}}

\definecolor{medium-green}{rgb}{0,0.6,0}

\journal{Computers and Mathematics with Applications}

\begin{document}

\begin{frontmatter}



\title{Supermodeling, a convergent data assimilation meta-procedure used in simulation of tumor progression}


\author[label1]{Maciej Paszy\'{n}ski}
\author[label1]{Leszek Siwik}
\author[label1]{Witold Dzwinel}
\author[label2]{Keshav Pingali}
\address[label1]{AGH University of Sciences and Technology \\ Faculty of Computer Science, Electronics and Telecommunication \\ Department of Computer Science \\ al. A Mickiewicza 30, 30-059 Krakow, Poland \\ email: maciej.paszynski@agh.edu.pl, phone: +48-12-328-3314, fax:  +48-12-617-5172 }
\address[label2]{The University of Texas at Austin \\ Institute for Computational and Engineering Sciences}

\begin{abstract}
Supermodeling is a modern, model-ensembling paradigm that integrates several self-synchronized imperfect sub-models by controlling a few meta-parameters to generate more accurate predictions of complex systems' dynamics. Continual synchronization between sub-models allows for trajectory predictions with superior accuracy compared to a single model or a classical ensemble of independent models whose decision fusion is based on the majority voting or averaging the outcomes. However, it comes out from numerous observations that the supermodeling procedure's convergence depends on a few principal factors such as (1) the number of sub-models, (2) their proper selection, and (3) the choice of the convergent optimization procedure, which assimilates the supermodel meta-parameters to data. Herein, we focus on modeling the evolution of the system described by a set of PDEs. We prove that supermodeling is conditionally convergent to a fixed-point attractor regarding only the supermodel meta-parameters. In our proof, we assume constant parametrization of the sub-models. We investigate the formal conditions of the convergence of the supermodeling scheme theoretically. We employ the Banach fixed point theorem for the supermodeling correction operator, updating the synchronization constants' values iteratively. From the theoretical estimate, we make the following conclusions. The ``nudging'' of the supermodel to the \textit{ground truth} (real data assimilated) should be well ballanced because both too small and too large attraction to data cause the supermodel desynchronization. The time-step size can control the convergence of the training procedure, by balancing the Lipshitz continuity constant of the PDE operator. All the sub-models have to be close to the ground-truth along the training trajectory but still sufficiently diverse to explore the phase space better. As an example, we discuss the three-dimensional supermodel of tumor evolution to demonstrate the supermodel's perfect fit to artificial data generated based on real medical images.
\end{abstract}

\end{frontmatter}



\section{Introduction}

It is well known that the multi-model ensembling approach can be a closer metaphor of an observed phenomenon than a single-model forecast~\cite{MultimodelWeigl2008}. However, averaging trajectories from multiple models without synchronization leads to undesired smoothing and variance reduction \cite{MultimodelHazeleger2015}. The alternative approach for taking advantage of many trajectories followed by distinctive models and discovering many "basin of attractions" without (premature) loss of the trajectories diversity is combining models dynamically. One of the first naive approaches of this kind has been proposed in \cite{Kirtman2003}, while more mature ones can be followed in \cite{SeltenDuane2017,Wu2013}. They introduce connection terms into the model equations that $nudge$ the state of one model to each other's states in the ensemble.

The computer model's assimilation to a real phenomenon through a set of observations is a complex inverse problem; thus, its time complexity increases exponentially with the number of parameters. This makes data assimilation procedures useless when applied for multiscale models such as weather/climate forecast or complex biological processes like tumor evolution \cite{SeltenDuane2017,Wiegerinck2013,isotumor2D,isotumor3D}.
Supermodel, the ensemble of dynamically synchronized sub-models, was applied in weather/climate forecast \cite{supermodel1,supermodel2,supermodel3,c19,c20}, for simulations of geological processes \cite{PhysRevLett.86.4298}, atmospheric phenomena \cite{Duane98co-occurrenceof,Duane98co-occurrenceof2}, and for the tumor evolution simulations \cite{DZWINEL20171832,DZWINEL2016999}.

Supermodel consisting of the sub-models represented by the same baseline model parametrized with different parameter sets has been analyzed in \cite{Sendera2020,Kocarev}. The authors show that coupling between them can be defined by a radically smaller number of meta-parameters than the number of parameters used in the baseline model. So, instead of matching tens parameters of a single model to observed data, just a few meta-parameters of the supermodel can be sufficient, while the original ones in the sub-models remain constant. These constant parameters can be initialized by an expert or can be generated by diverse solutions of fast pre-training.

In this paper, we investigate the conditions of the supermodeling scheme's convergence theoretically and approve them experimentally employing the supermodel of 3D tumor dynamics. Our principal contributions are: (1) providing mathematical proof of conditional convergence of the supermodel to an approximate trajectory assimilated to data and (2) developing a complex supermodel of tumor growth in 3D with perfect data assimilation ability.

The structure of the paper is the following. Section 2 introduces the supermodeling procedure. Section 3 uses the Banach fixed-point theorem for analysis of the supermodeling procedure, and draws some practical conclusions on the construction of the supermodel. Next, Section 4 presents the example of the supermodel construction concerning the tumor progression simulations with isogeometric analysis solvers. Finally, Section 5 discusses the numerical results on the tumor supermodel and their relation to the Banach fixed point theorem. We conclude the paper in Section 6.

\section{Supermodeling procedure}

The supermodel is defined as an ensemble of $N$ imperfect sub-models ${\cal F}_i, i=1,...,N$
synchronized with each other and with observed \emph{ground-truth} (GT) data ${\cal B}$ \cite{Sendera2020}. 

We focus on an evolution of a system $[0,T]\ni t \rightarrow B(t)\in {\cal B}$,where ${\cal B}$ is the set of all possible states of the system in a particular time moment, and we introduce time moments $0=t_0<t_1<t_2<\cdots<t_M=T$. Let us define $N$ sub-models, each of them able to mimic the system's evolution independently. Denote by
$B_i^n \in {\cal B}$ the state of sub-model $i=1,...,N$ at the time moment $t_n$. We assume that the sub-models are the baseline model's various instances, e.g., defined by its various parametrization.
Let assume that ${\cal F}^n: {\cal B}^N \times {\cal R}^{N^2} \ni (B^n_1,...,B^n_N) \times (C_{11},...,C_{NN})\rightarrow $ $\left({\cal F}^n_1(B^n_1,...,B^n_N;C_{11},...,C_{NN}),\cdots, \right.$ $\left. {\cal F}^n_N(B^n_1,...,B^n_N;C_{11},...,C_{NN})\right)$ $\times (C_{11},...,C_{NN})
 \in {\cal B}^N\times {\cal R}^{N^2}$:

\begin{eqnarray}
\begin{aligned}
B_i^{n+1}={\cal F}^n_i(B^n_1,...,B^n_N;C_{11},...,C_{NN})= \nonumber \\
 B_i^n+dt {\cal S} (B_i^n, {\cal P}_i)+ 
\inblue{\frac{1}{N}}\sum_{j=1,...,N}C_{ij} (B_j^n-B_i^n)+\inblue{\frac{1}{N}}K \sum_{j=1,...,N} (GT^n-B_j^n)
\end{aligned}\end{eqnarray}
where ${\cal S}(B_i^n,{\cal P}_i)$ denotes the solution of the PDEs, which model the system evolution at a time moment $t_n$ for the set of model parameters ${\cal P}_i$. Here $C_{ij}$ are the synchronization constants, and $K$ is the \emph{nudging} parameter responsible for attracting the supermodel to the measured GT data.
The sub-models differs in parameter sets ${\cal P}_i$ and are coupled with each other via dynamic variables ${\cal B}_i$ through the ``strings'' $({\cal B}^n_i - {\cal B}^n_j)$ parameterized by coupling factors $C_{ij}$.

The supermodel output is defined as the average outputs of the synchronized sub-models.
${\cal F_S}^n: {\cal B}^N \times {\cal R}^{N^2} \ni (B^n_1,...,B^n_N;C_{11},...,C_{NN}) \rightarrow $ ${\cal F_S}^n(B^n_1,...,B^n_N;C_{11},...,C_{NN})=\frac{1}{N}\sum_{i=1,...,N}{\cal F}^n_i(B^n_1,...,B^n_N;C_{11},...,C_{NN}) \in {\cal B}$.

We focus on the example of the supermodeling of tumor, where the state of the system is described by the scalar tumor density field $\Omega \times [0,T] \ni (x,y,z) \times t \rightarrow B(x,y,z;t)\in R$.
The supermodeling consists of the following phases:

\begin{enumerate}
\item {\bf Initialization}, we look for $N$ sets of parameter*s ${\cal P}_1,\cdots,{\cal P}_N$ to instantiate $N$ different models of tumor evolution.
\begin{itemize}
\item Select randomly $N$ different sets of model parameters ${\cal P}_1,\cdots,{\cal P}_N$ from pre-defined and physically justified intervals. The baseline parameters can be set-up by an expert or possible solutions of fast pre-testing of the model using classical DA algorithms. The sub-models can be sufficiently diverse to explore better the phase space. Sub-models diversity is the basic assumption in developing any ensemble classifiers (e.g., in machine learning).
\item It is reasonable to perform sensitivity analysis before the data assimilation process to find the most sensitive parameters and dynamic variables of the tumor model. Afterward, we can select a better - more diversified - set of the sub-models.
\item Once we have found $N$ sets of parameters, we can instantiate $N$ sub-models \emph{sim1}, ...., \emph{simN}, resulting in different tumor progressions such as those which snapshots are shown in Figure \ref{fig:3sims}.
\begin{figure}[h]
\begin{center}
\includegraphics[width=0.3\textwidth]{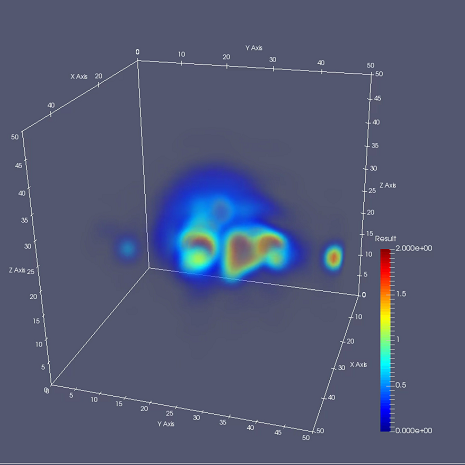}
\includegraphics[width=0.3\textwidth]{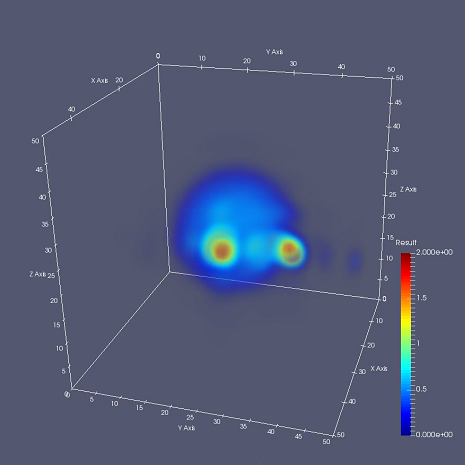}
\includegraphics[width=0.3\textwidth]{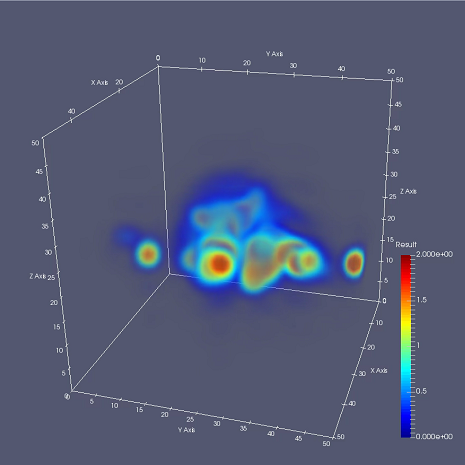}
\end{center}
\caption{Tumor evolution simulations within three diverse sub-models.}
\label{fig:3sims}
\end{figure}
\end{itemize}
\item {\bf Training} - finding the proper values of the coefficients $C_{ij}$ (supermodel meta-parameters) coupling the tumor density fields from $N$ sub-models.
\begin{itemize}
\item We select the most sensitive dynamic variables and define the coupling topology having in mind that the number of couplings can be small compared to the number of original model parameters.
\item Setup initial values of the coupling weights, e.g., $C_{ij}$ for $i,j=1,...,N$,$i \neq j$, select initial value of the $K$ coefficient, ``nudging'' the supermodel to the GT. The suggestions of the starting $C_{ij}$ values are discussed in \cite{Kocarev}. 
\item {\bf Iterate} with time steps until convergence or the maximum number of time steps has been reached.
\begin{enumerate}
\item We run a one-time step of each of the $N$ sub-models implemented in the simulators ($sim1$, ..., $simN$)
\item We pick up the tumor volume values from the GT data in a current time step. In case of insufficient data size, make GT data augmentation (approximation) to avoid underfitting. On the other hand, beware of overfitting.
\item We correct the tumor density fields as estimated by the $N$ sub-models, using the coupling factors $C_{ij}$ and $nudging$ coefficient $K$.
\begin{eqnarray}
\begin{aligned}
& B_i(x,y,z,t) +=  \sum_{j=1,...,N;i\neq j} \\ & \inblue{\frac{1}{N}} C_{ij} \left(B_j(x,y,z,t)-B_i(x,y,z,t)\right)+ \\ &
\inblue{\frac{1}{N}}K\left(GT(x,y,z,t)-B_j(x,y,z,t)\right) 
\end{aligned}
\end{eqnarray} where $GT(x,y,z)s$ represents the GT.
\item Correct the coupling parameters
\begin{eqnarray}
C_{ij}=A C_{ij}+ 
A\int_{\Omega} \left(GT(x,y,z,t)-B_i(x,y,z,t)\right)\\
 \left(B_i(x,y,z,t)-B_j(x,y,z,t)\right)dxdydz  \nonumber
\end{eqnarray}
\end{enumerate}
\end{itemize}
\item {\bf Supermodel simulation}. Once we have the coupling parameters trained, we can simulate with the supermodel in the prediction mode. We proceed the same way as in the training phase, but we use the coefficients' values obtained during the training. The following steps are as follows:
\begin{itemize}
\item Setup identical initial conditions in each of sub-models.
\item Use coupling weights $C^b_{ij}$ for $i,j=1,...,N$, and $K$ coefficient as obtained in the training procedure.
\item {\bf Iterate} with time-steps of the supermodel simulation
\begin{enumerate}
\item Run one time-step of each sub-model simulator ($sim1$, ..., $simN$)
\item Modify the obtained scalar fields in each of the sub-model using the trained coupling constants, i.e.,
\begin{eqnarray}
\begin{aligned}
& B_i(x,y,z,t) +=  \sum_{j=1,...,N;i\neq j}\\ & \inblue{\frac{1}{N}}C_{ij} \left(B_j(x,y,z,t)-B_i(x,y,z,t)\right)+ \\ & 
\inblue{\frac{1}{N}}K\left(GT(x,y,z,t)-B_j(x,y,z,t)\right) 
\end{aligned}
\end{eqnarray}
\item Average the sub-model scalar fields to calculate the supermodel output
\begin{eqnarray}
{\cal F}_S(x,y,z,t) = \sum_{i=1,...,N} \frac{1}{N} B_i(x,y,z,t) 
\end{eqnarray}
\end{enumerate}
\end{itemize}
\end{enumerate}

\section{Convergence of the supermodeling algorithm}



Let us assume that 

$$
B_i^{n+1}=
{\cal F}^n_i(B_1^n,\cdots,B_N^n;C_{11},...,C_{NN})=
$$
$$
 \inred{B_i^n+dt*{\cal S}(B_i^n,P_i)}+
 \inblue{\frac{1}{N}}\sum_{j=1,...,N}\left(C_{ij} (B_j^n-B_i^n)+
  K  (GT^n-B_j^n)\right)
$$
for $i=1,...,N$. The states of $i=1,...,N$ sub-models in time step $t_n$ are updated step-by-step by ${\cal F}^n_i$ operators 
$$
(B_1^{n+1},...,B_N^{n+1};C_{11},...,C_{NN}) = 
\left({\cal F}^n_1 (B_1^n,...,B_N^n;C_{11},...,C_{NN}),...,\right.
$$
$$
\left.{\cal F}^n_N (B_1^n,...,B_N^n;C_{11},...,C_{NN});C_{11},...,C_{NN}\right)=
$$
$$
\left({\cal F}^n (B_1^n,...,B_N^n;C_{11},...,C_{NN});C_{11},...,C_{NN}\right).
$$

Let us define the supermodeling correction operator. It corrects the supermodeling algorithm 
by adjusting the coupling constants $C_{ij}$

$$
{\cal G} \left( (B_1^{1,k},...,B_N^{1,k}),..., (B_1^{M,k},...,B_N^{M,k}); C_{11}^{k},...,C_{NN}^{k}\right)=
$$
$$
\left((B_1^{1,k+1},...,B_N^{1,k+1}),...,(B_1^{M,k+1},...,B_N^{M,k+1});C_{11}^{k+1},...,C_{NN}^{k+1}\right)
$$

where
$$
\left(B_1^{n+1,k+1},...,B_N^{n+1,k+1}\right)=
$$
$$
\left( {\cal F}^n_1(B_1^{n,k},\cdots,B_N^{n,k};C_{11}^{k+1},...,C_{NN}^{k+1}), \cdots , {\cal F}^n_N(B_1^{n,k},\cdots,B_N^{n,k};C_{11}^{k+1},...,C_{NN}^{k+1})\right)
$$
and
\begin{equation}
C_{ij}^{k+1}=AC_{ij}^{k}+A\max_{n=1,...,M}(GT^n-B_i^{n,k},B_i^{n,k}-B_j^{n,k})
\label{eq:cij}
\end{equation}

In the $B^{n,k}_i$ terms, $i$ denotes the sub-model index, $n$ denotes the time step index, and $k$ denotes the supermodel version (with $k$-th set of coupling meta-parameters $C^k_{ij}$).
$GT^n$ denotes the ground-truth (the measurement) at time moment $t_n$.
In other words, the ${\cal G}$ operator takes current values of the coupling coefficients, runs the supermodel simulation for $M$ time steps, and generates trajectories of $N$ sub-models. Next, it corrects the coupling coefficients ($C_{ij}^k$ into $C_{ij}^{k+1}$), comparing the sub-models with themselves and concerning the GT (e.g., medical imaging data). It runs the supermodel simulations with new coupling coefficients. In this state, the supermodel is as close as possible to the GT data.

{\bf Lemma 1} 

The supermodeling correction operator ${\cal G}$ for $N$ sub-models is contracting

\begin{eqnarray}
||{\cal G}\left( B^k;C^k\right)-{\cal G}\left(D^k;E^k \right)|| \leq const_1 || B^k-D^k||  + const_2 || C^k-E^k||
\end{eqnarray}
 where 
 $$
 (B^k;C^k)=\left((B_1^{1,k},...,B_N^{1,k}),...,(B_1^{M,k},...,B_N^{M,k}); C_{11}^{k},...,C_{NN}^k\right)
 $$ 
 and 
 $$
( D^k;E^k)=\left( (D_1^{1,k},...,D_N^{1,k}),..., (D_1^{M,k},...,D_N^{M,k});E_{11}^{k},...,E_{NN}^{k}\right).
 $$
and $const_i<1$.


{\bf Proof.}We employ the taxi norm for each set of states, and maximum norm for a sequence of time steps, {\color{magenta} and maximum norm for the synchronization coefficients}
$$
||{\cal G}\left( B^{k};C^{k}\right)-{\cal G}\left(D^{k};E^{k} \right)||=
$$
$$
\max_{n=1,...,M}\sum_{i=1,...,N}
||{\cal F}^n_i(B_{1}^{n,k},...,B_{N}^{n,k};C_{11}^{k+1},...,C_{NN}^{k+1})-
$$
$$
{\cal F}^n_i(D_{1}^{n,k},...,D_{N}^{n,k};E_{11}^{k+1},...,E_{NN}^{k+1})||{\color{magenta}}
{\color{magenta} +\max_{C_{11},...,C_{NN}}|C_{ij}^{k+1}-E_{ij}^{k+1}|}=
$$
$$
\max_{n=1,...,M}\sum_{i=1,...,N}
||
\left( \inred{B_{i}^{n,k}+dt*{\cal S}(B_{i}^{n,k},P_i)}+\right.
 $$
 $$
\left. \inblue{\frac{1}{N}}\sum_{j=1,...,N}\left(C^{k+1}_{ij} (B_{j}^{n,k}-B_{i}^{n,k})+
  K  (GT^n-B_{j}^{n,k})\right)-\right.
 $$
 $$
\left. \left(\inred{D_{i}^{n,k}+dt*{\cal S}(D_{i}^{n,k},P_i)}\right)-\right.
 $$
 $$
\left. \inblue{\frac{1}{N}}\sum_{j=1,...,N}\left(E_{ij}^{k+1} (D_{j}^{n,k}-D_{i}^{n,k})+
  K  (GT^n-D_{j}^{n,k})\right) \right. ||{\color{magenta}}
{\color{magenta} +||C^{k+1}-E^{k+1}||}=
$$
 \begin{eqnarray}
\max_{k=1,...,M} \sum_{i=1,...,N} ||\left( (B_{i}^{n,k}-D_{i}^{n,k})+
 dt*{\cal S}(B_{i}^{n,k}-D_{i}^{n,k},P_i)+ \right. \nonumber\\
{\color{red} {\frac{1}{N}}\sum_{j=1,...,N} 
\left( \left(C^{k+1}_{ij}B_{j}^{n,k}-E^{k+1}_{ij}D_{j}^{n,k}\right) +
\left(E^{k+1}_{ij}D_{i}^{n,k}-C^{k+1}_{ij}B_{i}^{n,k}\right) \right)
}
 \nonumber\\
\left.-\inblue{\frac{1}{N}}\sum_{j=1,...,N}K  (B_{j}^{n,k}-D_{j}^{n,k})\right)|| 
{\color{magenta}+ ||C^{k+1}-E^{k+1}||}\leq \nonumber \\
  (a)+(c) {\left.\color{magenta}+ (b) \right)}
\label{eq:all}
\end{eqnarray}
  
Now, we consider three terms: 
(a) the red term following the update of the synchronization coefficients, 
 \begin{eqnarray}
(a)=  \max_{k=1,...,M} \sum_{i=1,...,N}  \nonumber 
 \end{eqnarray}
 \begin{eqnarray}
||
{\color{red} {\frac{1}{N}}\sum_{j=1,...,N} 
\left( \left(C^{k+1}_{ij}B_{j}^{n,k}-E^{k+1}_{ij}D_{j}^{n,k}\right) +
\left(E^{k+1}_{ij}D_{i}^{n,k}-C^{k+1}_{ij}B_{i}^{n,k}\right) \right)
}||
\label{eq:a}
\end{eqnarray}
(b) the magenta term following the maximum norm of the synchronization coefficients, 
 \begin{eqnarray}
(b)= {\color{magenta} ||C^{k+1}-E^{k+1}||}
\label{eq:b}
\end{eqnarray}
and (c) the remaining terms
 \begin{eqnarray}
(c) =
\max_{k=1,...,M} \sum_{i=1,...,N} ||\left( (B_{i}^{n,k}-D_{i}^{n,k})+
 dt*{\cal S}(B_{i}^{n,k}-D_{i}^{n,k},P_i)+ \right. \nonumber\\
\left.-\inblue{\frac{1}{N}}\sum_{j=1,...,N}K  (B_{j}^{n,k}-D_{j}^{n,k})\right)|| 
\label{eq:c}
\end{eqnarray}

The term (a) from equation (\ref{eq:all}) describing the update of the synchronization coefficients can be rewritten as

$$
{\color{red}\max_{n=1,...,M}\sum_{i=1,...,N}||\frac{1}{N}\sum_{j=1,...,N}}
{\color{red} \left(C^{k+1}_{ij}B_{j}^{n,k}-E^{k+1}_{ij}D_{j}^{n,k}\right)}+
{\color{red}\left(E^{k+1}_{ij}D_{i}^{n,k}-C^{k+1}_{ij}B_{i}^{n,k}\right)|| }
$$
$$
\leq \max_{n=1,...,M}\sum_{i=1,...,N}||\frac{1}{N}\sum_{j=1,...,N}
 \left(C^{k}_{ij}B_{j}^{n,k}-E^{k}_{ij}D_{j}^{n,k}\right)+
 \left(E^k_{ij}D_{i}^{n,k}-C^k_{ij}B_{i}^{n,k}\right)||
$$
$$
{\color{blue}+\max_{n=1,...,M}\sum_{i=1,...,N}||\frac{1}{N}\sum_{j=1,...,N}}
{\color{blue} \left(\max_{n=1,...,M} (GT^n-B_i^{n,k},B_i^{n,k}-B_j^{n,k}) B_j^{n,k} - \right.}
$$
$$
{\color{blue} \left. \max_{n=1,...,M}(GT^n-D_i^{n,k},D_i^{n,k}-D_j^{n,k}) D_j^{n,k}\right.+}
{\color{blue}  \max_{n=1,...,M}(GT^n-D_i^{n,k},D_i^{n,k}-D_j^{n,k}) D_i^{n,k}-}
$$
$$
{\color{blue} \left. \max_{n=1,...,M}(GT^n-B_i^{n,k},B_i^{n,k}-B_j^{n,k}) B_i^{n,k} \right)||}  = (d) + \inblue{(e)}
$$

The blue term \inblue{(e)} is bounded by
$$
{\color{blue} (e) \leq \max_{n=1,...,M}\sum_{i=1,...,N}||\frac{1}{N}\sum_{j=1,...,N}}
{\color{blue} \max \left(  \max_{i,j,n}  |(GT^n-B_i^{n,k},B_i^{n,k}-B_j^{n,k}) | ,\right.}
$$
$$
{\color{blue}\left.   \max_{i,j,n} |(GT^n-D_i^{n,k},D_i^{n,k}-D_j^{n,k})| \right) }
$$
$$
+{\color{blue}\max_{n=1,...,M}\sum_{i=1,...,N}  \frac{1}{N}}
{\color{blue}||   \sum_{j=1,...,N}(B_j^{n,k} -B_i^{n,k}-D_j^{n,k}+D_i^{n,k}) ||} 
$$

After taking into account averaged double summation outside the norm, 
and adding and subtracting GT to the first terms, we get

$$
{\color{blue}\max \left(   \max_{i=1,...,N;n=1,...,M} 2 ||GT^n-B_i^{n,k}||^2 , \max_{i=1,...,N;n=1,...,M} 2||GT^n-D_i^{n,k}||^2 \right) }
$$
$$
+{\color{blue}2\max_{n=1,...,M}\sum_{i=1,...,N} 
|| D_i^{n,k}-B_i^{n,k}|| = \beta ||B^k-D^k||}
$$

\begin{equation}
\beta = 
4\max \left(  ||GT-B^k||^2 , ||GT-D^k||^2 \right) \label{eq:beta}
\end{equation}

The term (d) 
$$
(d) \leq  
2\max_{i,j=1,...,N}\{ |C_{ij}^k-E_{ij}^k|\}
\left(\sum_{j=1,...,N}( B_j^{n,k}-D_j^{n,k})\right)
$$

Now we focus on term (c) in (\ref{eq:all}). We assume that our PDE operator is bounded ${\cal S}(B_i^{n,k}-D_i^{n,k},{\cal P}_i)\leq const({\cal S},{\cal P}_i)||B_i^{n,k}-D_i^{n,k}||$, where $const({\cal S},{\cal P}_i)$ is the Lipschitz continuity constant. We use the definition of the taxi norm to get

 $$
 \max_{n=1,...,M}\inred{\left((1-K)+dt*\max_i const({\cal S},{\cal P}_i)\right)}
\left(\sum_{i=1,...,N}(B_i^{n,k}-D_i^{n,k})\right)
$$

We group parts in terms (c)+(d)
$$
 \left(\inred{\left((1-K)+dt*\max_i const({\cal S},{\cal P}_i)\right)}+2max\{ |C^k_{ij}-E^k_{ij}|\}\right)
  \left(B^{k}-D^k\right) = 
$$
$$
\alpha|| B^k-D^k||
$$
\begin{eqnarray}
\alpha=
 \left(\inred{\left((1-K)+dt*\max_i const({\cal S},{\cal P}_i)\right)}+ 2max\{ |C^k_{ij}-E^k_{ij}|\}\right) \label{eq:alpha}
\end{eqnarray}

We focus now on (b) term
$$
{\color{magenta} ||C^{k+1}-E^{k+1}|| =\max_{i,j=1,...,N} |C^{k+1}_{ij}-E^{k+1}_{ij}| =}
$$
$$
{\color{magenta} A\max_{i,j=1,...,N}|C_{ij}^{k}+\max_{n=1,...,M}(GT^n-B_i^{n,k},B_i^{n,k}-B_j^{n,k})-}
$$
$$
{\color{magenta} E_{ij}^{k}-\max_{n=1,...,M}(GT^n-D_i^{n,k},D_i^{n,k}-D_j^{n,k})| =}
{\color{magenta} A\max_{i,j=1,...,N}|C_{ij}^{k}-E_{ij}^k|}+
$$
$$
{\color{magenta} +A\max_{n=1,...,M;i,j=1,...,N}}
{\color{magenta}|(GT^n-B_i^{n,k},B_i^{n,k}-GT^n+GT^n-B_j^{n,k})|}
$$
$$
{\color{magenta} +A\max_{n=1,...,M;i,j=1,...,N}}
{\color{magenta} |(GT^n-D_i^{n,k},D_i^{n,k}-GT^n+GT^n-D_j^{n,k})| \leq}
$$
$$
{\color{magenta} A||C^k-E^k||}
{\color{magenta} +2A||GT-B^k||^2}
{\color{magenta} +2A||GT-D^{k}||^2 }
{\color{magenta} = A||C^k-E^k|| + \gamma }
$$

\begin{equation}
\gamma = 2A||GT-B^k||^2 +2A||GT-D^{k}||^2 \label{eq:gamma}
\end{equation}

The whole operator
$$
||{\cal G}\left( B^{k};C^{k}\right)-{\cal G}\left(D^{k};E^{k}) \right)|| \leq
(\alpha+\beta) || B^k-D^k|| +  A||C^k-E^k||+\gamma
$$

The operator ${\cal G}$ is contracting, if $\alpha \leq 1 $, $A\leq 1$, and $\beta,\gamma \approx 0$, with $\alpha$ given by (\ref{eq:alpha}), $\beta$ by (\ref{eq:beta}) and $\gamma$ by (\ref{eq:gamma}).

{\bf Remark 1}
From the Banach fixed-point theorem \cite{Agarwal2018}, the contracting operators converge to the fixed-point. In the case of ${\cal G}$ operator, the fixed point means the supermodel trajectory that approximates the GT.

{\bf Remark 2}
To guarantee the convergence
\begin{itemize}
\item The nudging $K$ has to be close to 1.0 (see $\alpha$ (\ref{eq:alpha})). 
\item PDE operator ${\cal S}(B_i^{n,k}-D_i^{n,k},{\cal P}_i)$ needs to be linear and bounded
$||{\cal S}(B_i^{n,k}-D_i^{n,k},{\cal P}_i)|| \leq const({\cal S},{\cal P}_i)||B_i^{n,k}-D_i^{n,k}||$ (see $\alpha$ (\ref{eq:alpha})). We can control this norm by decreasing time-step size $dt$. However, in the case when the sub-models are too distant and instead of synchronization they start to diverge, by setting $dt=0$ we stop model training, what may signal the need of selecting a different ensemble.
\item Since $|a-b|\leq \max\{|a|,|b|\}$ (for $a,b\geq 0$), so the term $|C^k_{ij}-E^k_{ij}|\leq \max\{|C^k_{ij}|,|E^k_{ij}|\}$. We can control this term by assuming limits for the coupling coefficients values (see $\alpha$ (\ref{eq:alpha})).
\item All sub-models has to be close to the GT, meaning small values of $||GT-B^{k}||^2$ , $||GT-D^{k}||^2$ for the entire trajectory (see $\beta$ (\ref{eq:beta}) and $\gamma$ (\ref{eq:gamma})). \end{itemize}

{\bf Remark 3}
Assuming $\beta,\gamma \approx 0$, the rate of convergence is given by
$$
||{\cal G}\left( B^{k};C^{k}\right)-{\cal G}\left(D^{k};E^{k}) \right)|| \leq
\frac{A}{1-\alpha} ||C^k-E^k||.
$$
This means that if we disturb the parameters ${\cal P}_i$ of sub-models, the Lipschitz constants $const({\cal S},{\cal P}_i)$ inside $\alpha$ may change, and we may get different convergence rates, related to different values of $\alpha$.

\section{Supermodel of tumor}

\begin{figure}	
\centering
	\includegraphics[width=0.5\textwidth]{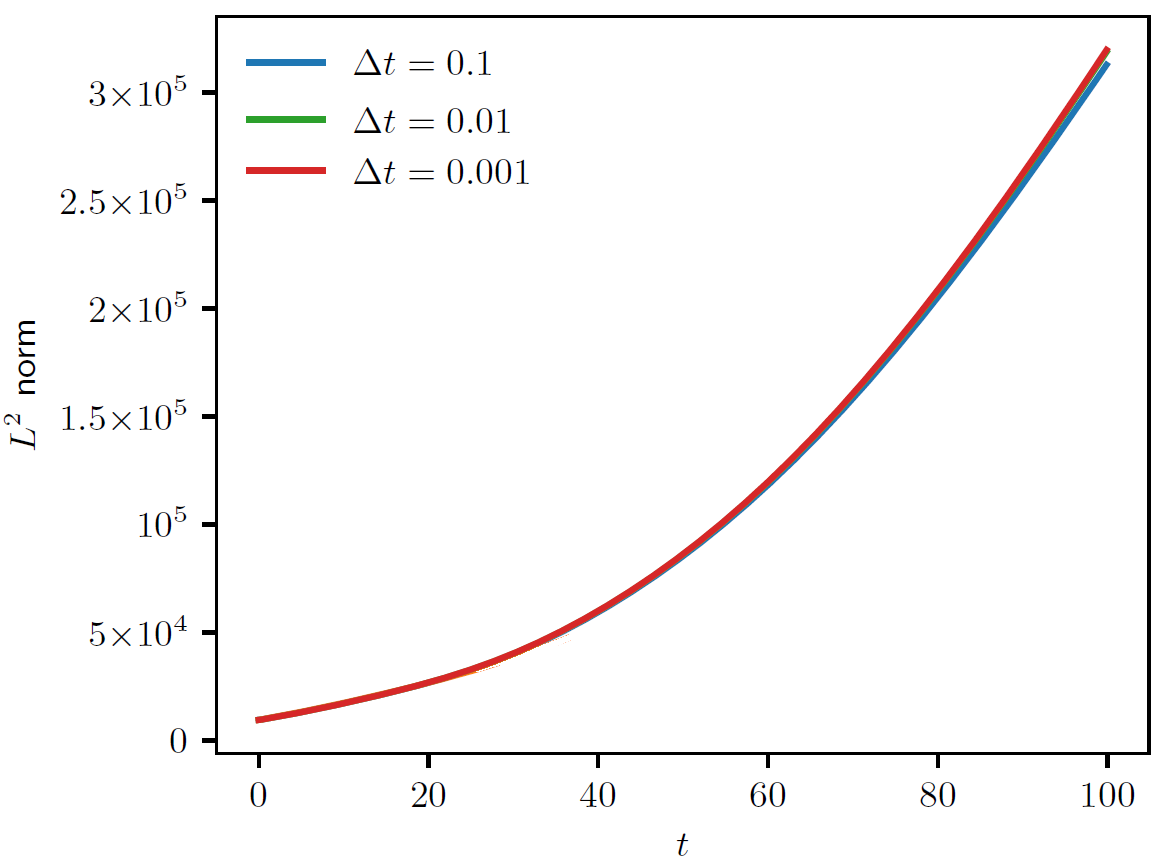}
	\caption{Verification of the CFL condition for $dt=0.1,0.01,0.001$ for mesh size $64\times 64\times 64$.}
	\label{fig:CFL}
\end{figure}

In this section, we introduce an exemplary challenging supermodel for three-dimensional simulations of the progression of tumor.
For the numerical simulations, we employ the tumor progression model described in \cite{isotumor3D}, using the isogeometric analysis solver.
To summarize, the model is described by the following set of mainly parabolic, diffusion-reaction type of PDEs equations:
\begin{equation}
	\left\{
	\begin{alignedat}{3}
		\frac{\partial \inred{b}}{\partial t} &= -\nabla \cdot \left(- D_b\, \inred{b} \left(\nabla (\frac{\inred{b} - b^N}{b^M - b^N})[b^N\leq \inred{b}\leq b^M] + r_b \nabla A\right)\right) - \\ & \quad \frac{\inred{b}}{\inblue{T^{death}}}[ o < \inblue{o^{death}}] + \\ & \quad \frac{\inred{b}}{\inblue{T^{prol}}}\left( 1 + \frac{\tau_b A}{\tau_b A + 1} P_b \right) \left( 1 - \frac{\inred{b}}{b^M} \right)[o > \inblue{o^{prol}}] \\
		\frac{\partial c}{\partial t}&=  \chi_c \Delta c - \gamma_c o c+ b (1 - c) [o < o^{death}]\\
		\frac{\partial o}{\partial t}  &=  \alpha_0 \Delta o -\gamma_o \inred{b} o +\delta_o \left(o^{max} - o\right)  \\
		\frac{\partial M}{\partial t} &=  -\beta_M M \inred{b}  \\
		\frac{\partial A}{\partial t}&=  \gamma_A M \inred{b} + \chi_{OA} \Delta A - \gamma_{OA} A 
	\end{alignedat} 
	\right. \label{eq:system}
\end{equation}
We have denoted in red the most sensitive dynamic variable (the tumor cell density scalar field $\inred{b}$) that we will use in coupling and synchronization of sub-models. We have denoted in blue the most sensitive model parameters, as found by the sensitivity analysis \cite{Siwik}. 
Additionally, the oxygen is generated along the dynamically updated graph of vessels, following the model described in \cite{VascularNetwork}.
Summarizing, the model is controlled by twenty-one parameters, presented in Table \ref{tab:parameters}.
\begin{table}
	\centering
	\scalebox{.7}[0.7]{
		\begin{tabular}{ | l | r | p{7cm} |}
			\hline
			\textbf{Symbol} & \textbf{Value} & \textbf{Description}            \\ \hline
			$b_m$        & 0         & min tumor cell density                  \\ \hline
			$b_M$        & 2         & max tumor cell density                  \\ \hline
			$b^{norm}$   & 1         & normal tumor cell density               \\ \hline
			$D_b$        & varies    & tumor cell diffusion rate               \\ \hline
			$r_b$        & $0.3$     & tumor cells chemoattractant sensitivity \\ \hline
			$\inblue{o^{prol}}$   & 10        & \inblue{tumor proliferation threshold}           \\ \hline
			$\inblue{o^{death}}$  & 2         & \inblue{tumor cell hypoxia threshold}            \\ \hline
			$\inblue{T^{prol}}$   & 10        & \inblue{tumor cell proliferation time}           \\ \hline
			$\inblue{T^{death}}$  & 100       & \inblue{tumor cell survival time}                \\ \hline
			$P_b$        & 0.001     & maximum stimulated mitosis rate         \\ \hline
			$\tau_b$     & 0.5       & instantaneous reaction rate             \\ \hline
			$\beta_M$    & 0.0625    & ECM decay rate                          \\ \hline
			$\gamma_A$   & 0.032     & production rate of attractants          \\ \hline
			$\chi_{aA}$  & 0.000641  & decay rate of digested ECM              \\ \hline
			$\gamma_{oA}$& 0.000641  & diffusion rate of digested ECM          \\ \hline
			$\chi_c$     & 0.0000555 & TAF diffusion rate                      \\ \hline
			$\gamma_c$   & 0.01      & TAF decay rate                          \\ \hline
			$\alpha_o$   & 0.0000555 & oxygen diffusion rate                   \\ \hline
			$\gamma_o$   & 0.01      & oxygen consumption rate                 \\ \hline
			$\delta_o$   & 0.4       & oxygen delivery rate                    \\ \hline
			$o^\text{max}$ & 60      & maximal oxygen concentration            \\ \hline
	\end{tabular}}
	\captionof{table}{Glossary of tumor model parameters \cite{isotumor3D}.}
	\label{tab:parameters}
\end{table}
Each of the tumor supermodel equation, as listed in (\ref{eq:system}) has the following terms:
\begin{itemize}
    \item Diffusion terms, e.g. $const* \Delta u$
    \item Reaction terms $const* u$
    \item Non-linear terms $const*\nabla \cdot (u \nabla u)$
    \item Mixed non-linear terms $const*\nabla \cdot (u \nabla w)$
    \item Mixed terms $uw$
\end{itemize}
where $u,w \in \{b,c,o,M,A,J\}$,
and $const$ are expressed as combination of the model parameters, summarized in Table \ref{tab:parameters}.
This non-linear terms are treated explicitly, and the higher-order derivatives are split by introduction of the \ingreen{auxiliary variables}. We split $const * \nabla \cdot (const*b \nabla b+const*b\nabla A)$ into $const * \nabla \cdot J$ and $J=(const b\nabla b+const b\nabla A)$.
In our simulation, we also update the graph vessels with time steps \cite{VascularNetwork}, and thus we cannot use too large time steps. 
Thus, the whole split system is treated explicitly as given by\begin{equation}
	\left\{
	\begin{alignedat}{3}
		\inred{b_t+1} &= b_t +dt \{-\nabla \cdot \ingreen{J_t} - \frac{\inred{b_t}}{\inblue{T^{death}}}[ o_t < \inblue{o^{death}}] + \\ & \quad \frac{\inred{b_t}}{\inblue{T^{prol}}}\left( 1 + \frac{\tau_b A_t}{\tau_b A_t + 1} P_b \right) \left( 1 - \frac{\inred{b_t}}{b^M} \right)[o_t > \inblue{o^{prol}}]\} \\
		\ingreen{J_{t}} & \ingreen{= - D_b\, b_t \left(\nabla \left((\frac{\inred{b_t} - b^N}{b^M - b^N})[b^N\leq \inred{b}\leq b^M]\right) + r_b \nabla A_t\right)} \\
		c_{t+1}&= c_t+dt\{ \chi_c \Delta c_t - \gamma_c o_t c_t+ b (1 - c_t) [o_t < o^{death}] \}\\
		o_{t+1}  &= o_t+dt\{ \alpha_0 \Delta o_t -\gamma_o \inred{b_t} o_t +\delta_o \left(o^{max} - o_t\right)\}  \\
		M_{t+1} &= M_t+dt\{ -\beta_M M_t \inred{b}\}  \\
		A_{t+1} &= A_t+dt\{ \gamma_A M_t \inred{b_t} + \chi_{OA} \Delta A_t - \gamma_{OA} A_t \}
	\end{alignedat} 
	\right. \label{eq:system2}
\end{equation}
We transform all equations into the weak forms
\begin{equation}
	\left\{
	\begin{alignedat}{3}
		\left(\inred{b_t+1},v\right) &= \left(b_t,v\right) +dt \{-\left(\nabla \cdot \ingreen{J_t},v\right) - \frac{\left(\inred{b_t},v\right)}{\inblue{T^{death}}}[ o_t < \inblue{o^{death}}] + \\ & \quad \left(\frac{\inred{b_t}}{\inblue{T^{prol}}}\left( 1 + \frac{\tau_b A_t}{\tau_b A_t + 1} P_b \right) \left( 1 - \frac{\inred{b_t}}{b^M} \right)[o_t > \inblue{o^{prol}}],v\right)\} \\
		\ingreen{\left(J_{t},v\right)} & \ingreen{= - D_b\, \left( b_t \nabla \left(\frac{\inred{b_t} - b^N}{b^M - b^N}\right)[b^N\leq \inred{b_t}\leq b^M],v\right) +} \\ & \quad  \ingreen{- D_br_b\, \left( b_t  \nabla A_t,v\right)} \\
	\left(c_{t+1},v\right)&= \left(c_t,v\right)+dt\{ -\chi_c \left(\nabla  c_t,\nabla v\right) - \gamma_c \left(o_t c_t,v\right) + \\ & \quad \left(b_t (1 - c_t) [o_t < o^{death}],v\right) \}\\
		\left(o_{t+1},v \right)  &= \left(o_t,v\right)+dt\{ -\alpha_0 \left(\nabla o_t,\nabla v\right) \\ & \quad -\gamma_o \left(\inred{b_t} o_t,v\right) +\delta_o \left(\left(o^{max} - o_t\right),v\right)\}  \\
		\left(M_{t+1},v\right) &= \left(M_t,v\right)-dt\beta_M\left( M_t \inred{b_t},v\right)\}  \\
		\left(A_{t+1},v\right) &= \left(A_t,v\right)+dt\{ \gamma_A \left(M_t \inred{b_t},v\right) - \chi_{OA} \left(\nabla A_t,v\right) - \gamma_{OA} \left(A_t,v\right) \}		
	\end{alignedat} 
	\right. \label{eq:system2}
\end{equation}
We utilize the B-spline basis functions for the approximation of all the scalar fields and averaging within the weak formulation. 
The explicit dynamics simulations require fulfillment of the CFL condition, namely to adjust the time step size according to the mesh size. So we experiment with different time step sizes and mesh dimensions, and we monitor the $L^2$ norm of the solution, as presented in Figure \ref{fig:CFL}. We select the time step $dt=0.1$ as providing numerically stable results on selected mesh dimensions $64\times 64\times 64$.
For efficient simulations of explicit dynamics with the isogeometric analysis we apply the IGA-ADS solver \cite{LOS201799}.

\begin{figure}
\includegraphics[width=0.8\textwidth]{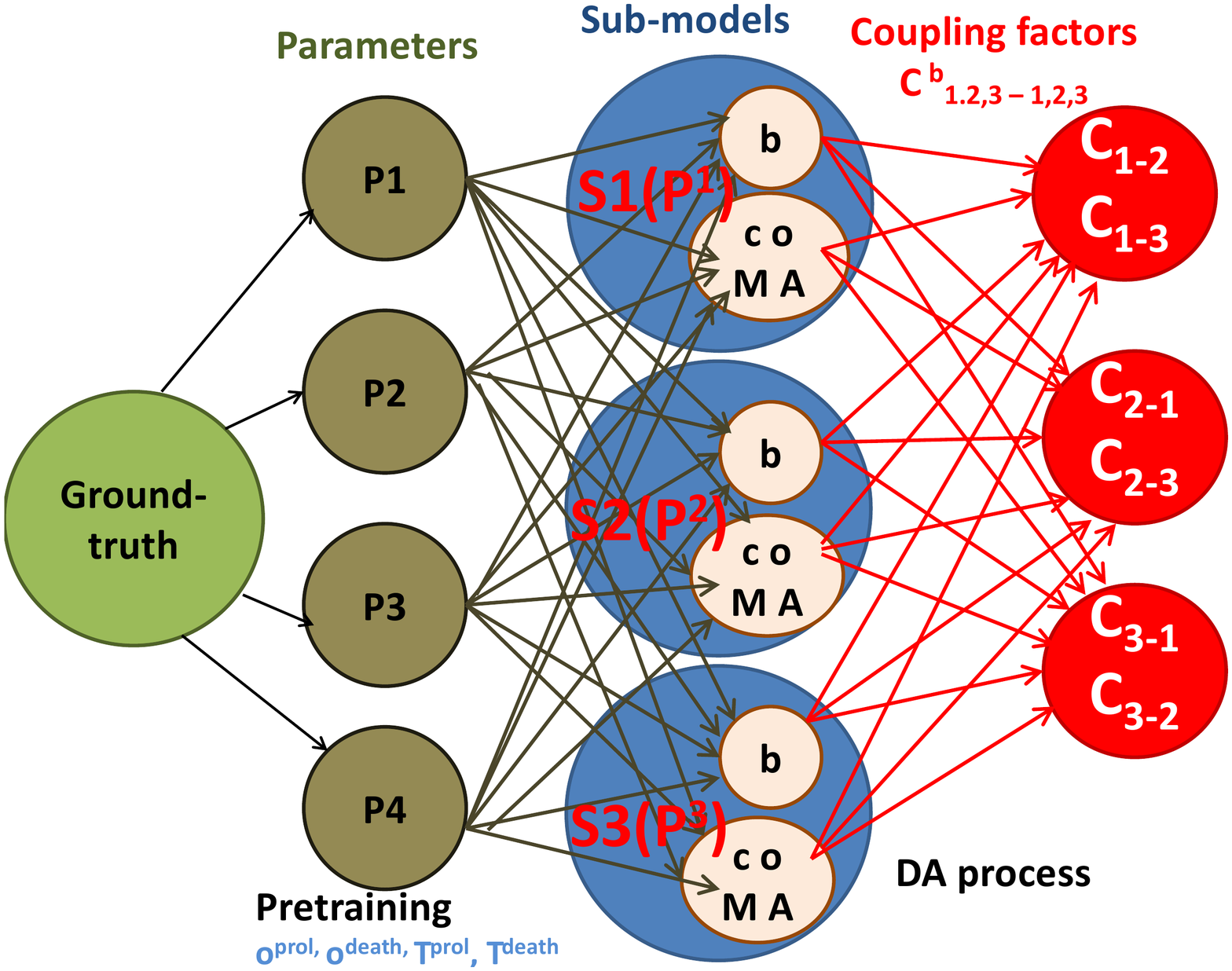}
\caption{Dynamic variable used for coupling: \inred{tumor cell density $b$} and most sensitive model parameters: \inblue{tumor cell proliferation threshold $o^{prol}$ and hypoxia threshold $o^{death}$,}  \inblue{tumor cell proliferation time $T^{prol}$ and survival time $T^{death}$}}
\label{fig:tumorsupermodel}
\end{figure}

The tumor supermodel construction, as shown in Figure \ref{fig:tumorsupermodel}, involves:
\begin{itemize}
\item Homogenous tumor PDEs model, with the isogeometric alternating directions solver \cite{isotumor3D} and the embedded dynamic discrete vasculature graph \cite{VascularNetwork}.
\item The sub-models are created with various sets of parameters representing different cancer growth scenarios.
\item The sub-models are coupled via only one dynamical variable - the tumor cell density $b(x,y,z,t)$.
\item As the ground-truth, we use artificial data of the tumor evolution approximated based on a few medical images.
\end{itemize}

\section{Numerical results}
To demonstrate the advantages of our approach, we present here a few numerical experiments of data adaptation to the supermodel of tumor.

\subsection{First experiment}
\label{sec:exp1}

In the first numerical experiment, we build three sub-models with randomly selected parameters that have been distributed as $\pm40\%$ around the reference parameters from Table 1. We vary the most sensitive model parameters, namely 
\inblue{tumor proliferation threshold}, \inblue{tumor cell hypoxia threshold},  \inblue{tumor cell proliferation time}, and \inblue{tumor cell survival time}. These sub-models correspond to different scenarios of tumor growth. We start the training phase with the coupling constants $C_{ij}= 0.5$, and the parameter coupling with the reality $K=2.0$. We have set up the range of the $C_{ij}$ values between $[0.1,0.9]$.

\begin{figure}[H]
\includegraphics[page=1]{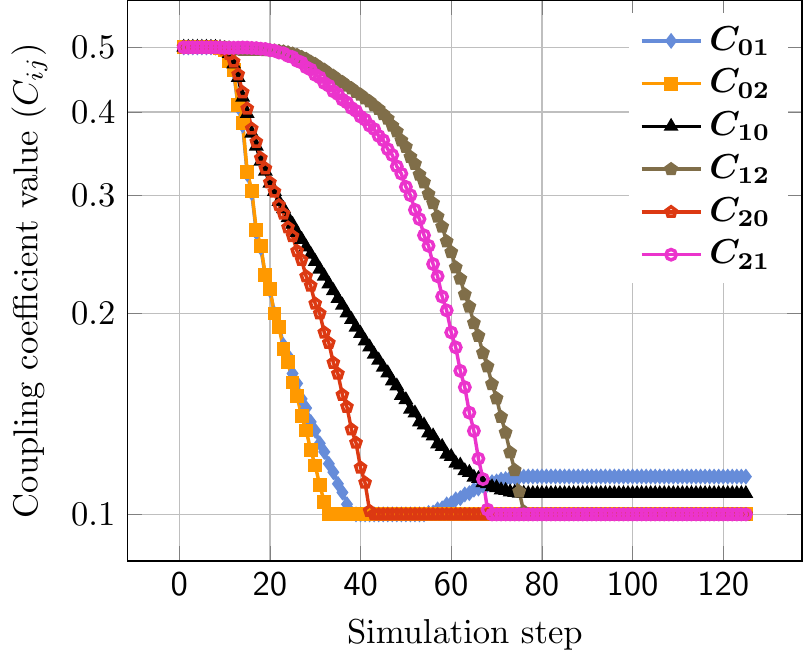}
\caption{Convergence of coupling coefficients $C_{ij}$ for ${K=2.0}$.}
\label{fig:sim1}
\end{figure}

In Figure \ref{fig:sim1} we present the dynamics of the coupling coefficients $C_{ij}$ during training. We can see that they do not converge since they reach the minimum possible value of $0.1$. The horizontal axis denotes the number of time steps, and the vertical axis indicates the $C_{ij}$ values.

\begin{figure}[H]
\includegraphics[page=1]{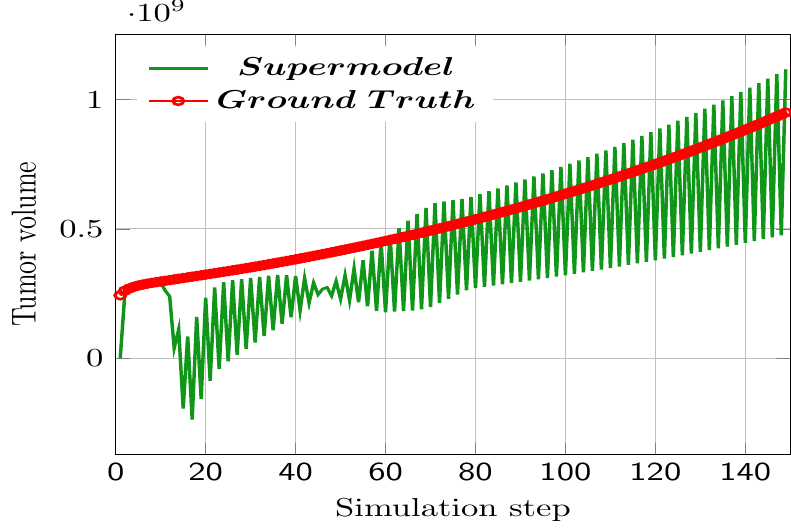}
  \caption{Convergence of tumor volumes for supermodel with respect to the GT.}
\label{fig:sim1a}
\end{figure}

\begin{figure}[H]
\includegraphics[page=1]{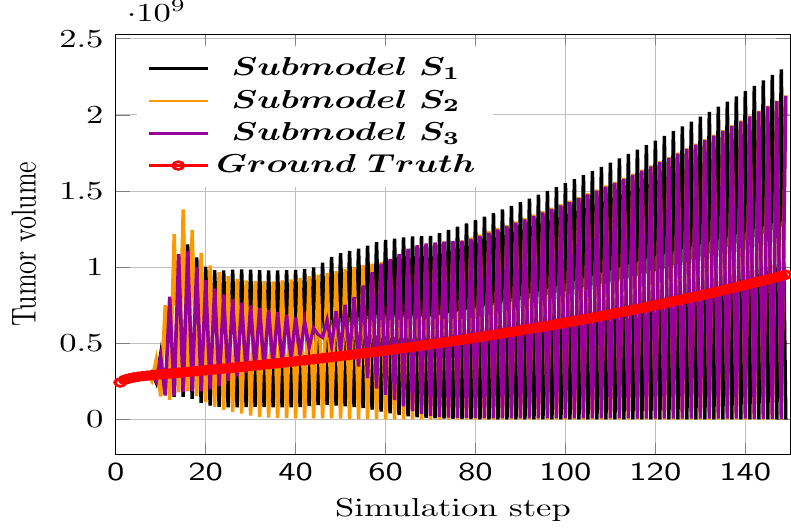}
  \caption{Convergence of tumor volumes for sub-models with respect to the GT.}
\label{fig:sim1b}
\end{figure}


We investigate the problem further by presenting in Figures \ref{fig:sim1a} and \ref{fig:sim1b} the supermodel's convergence to the GT, measured in terms of the total tumor volume. This time the horizontal axis denotes the number of time steps and the vertical axis denotes the total tumor volumes. We present the volumes for particular sub-models (\ref{fig:sim1a}), as well as for the average of the sub-models (\ref{fig:sim1b}) and the GT (''reality''). We can read from these figures that the sub-models and the supermodel fell into oscillations, and we find out that the reason is that our coefficient $K$ \emph{nudging} the supermodel to the reality is too big.

\begin{figure}[H]
\includegraphics[page=1]{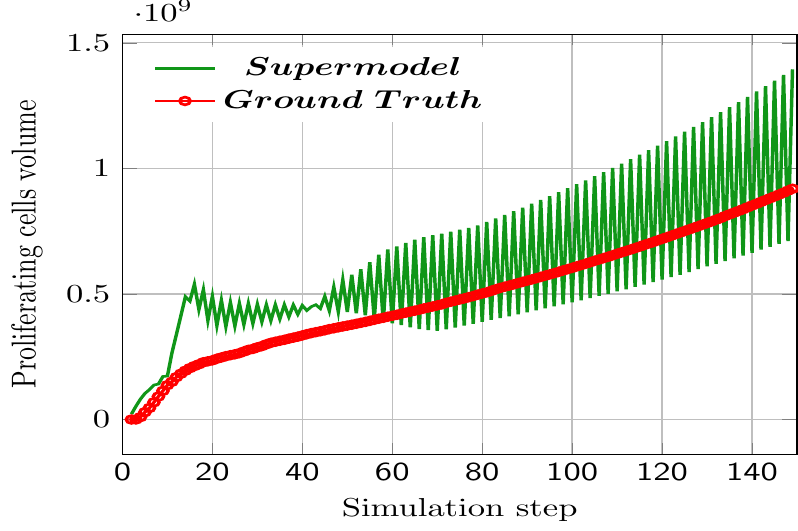}
  \caption{Convergence of the volumes of tumor proliferating cells for supermodel with respect to the GT.}
\label{fig:sim1c}
\end{figure}

\begin{figure}[H]
\includegraphics[page=1]{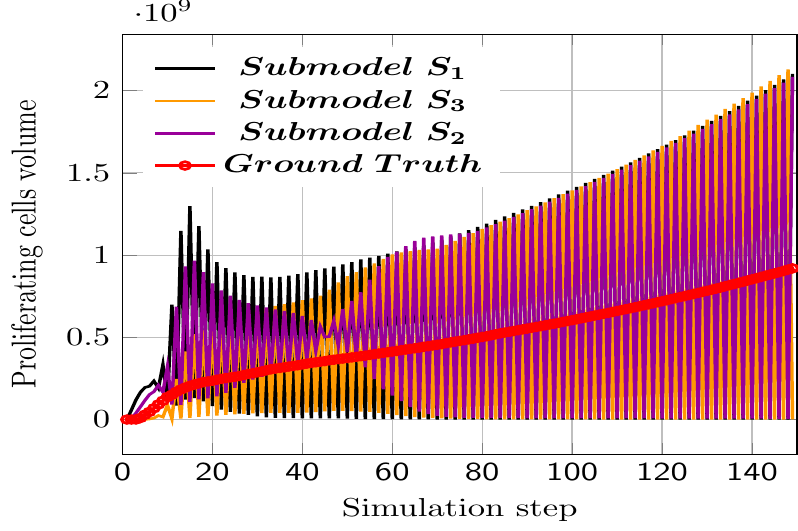}
  \caption{Convergence of the volumes of tumor proliferating cells for sub-models with respect to the GT.}
\label{fig:sim1d}
\end{figure}

Total tumor volume (presented in Figure \ref{fig:sim1a}) consists of the volume of proliferating and quiescent cells \citation{Ribba2012}.
Proliferating cells are those who multiply due to the mitosis in the presence of high local concentration of oxygen. The quiescent tumor cells are the cells that cannot proliferate due to unfavorable living conditions such as hypoxia or high external pressure (e.g., \cite{RIBBA2011479}).
Next, we present the supermodel's convergence to the GT, measured in terms of the proliferating tumor cells. As previously we present the volumes of proliferating cells for particular sub-models (\ref{fig:sim1d}), as well as for the average of the sub-models (\ref{fig:sim1c}) and the GT (''reality'').

\begin{figure}[H]
\includegraphics[page=1]{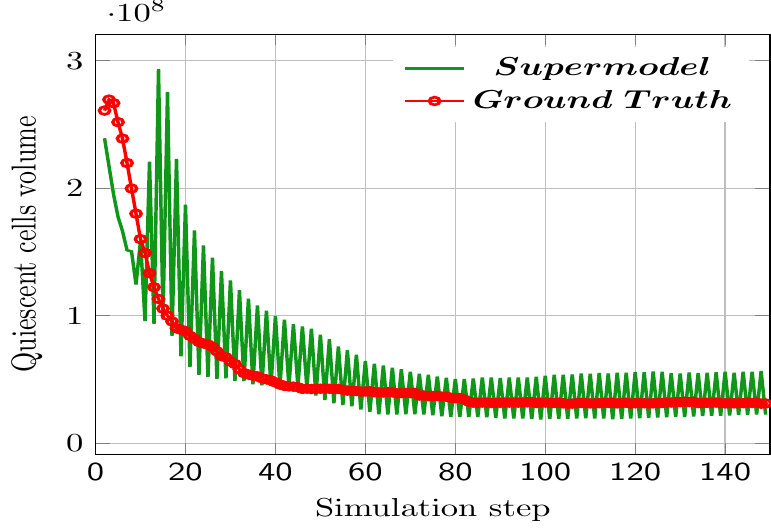}
  \caption{Convergence of the volumes of tumor quiescent cells for supermodel with respect to the GT.}
\label{fig:sim1e}
\end{figure}

\begin{figure}[H]
\includegraphics[page=1]{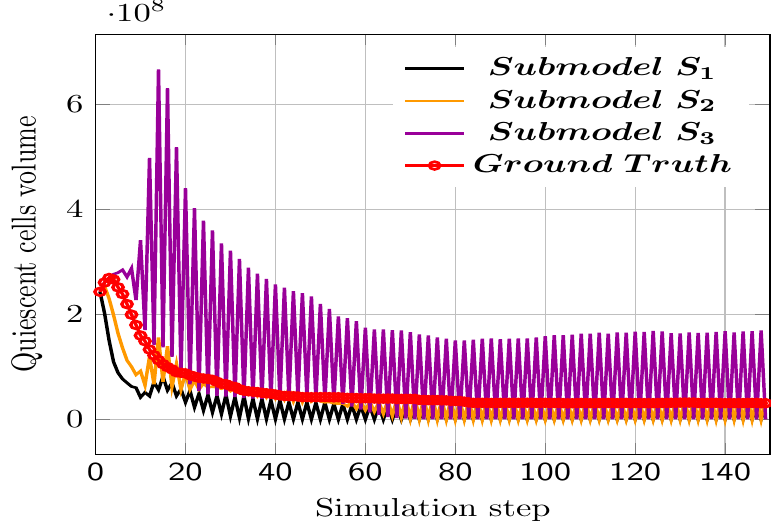}
  \caption{Convergence of the volumes of tumor quiescent cells for  sub-models  with respect to the GT.}
\label{fig:sim1f}
\end{figure}

Next, in Figures \ref{fig:sim1e} and \ref{fig:sim1f} we present the convergence of the supermodel to the GT, this time measured in terms of the quiescent tumor cells. We present the volumes of quiescent cells for particular sub-models (\ref{fig:sim1f}), for the average of the sub-models (\ref{fig:sim1e}), and for the GT.

\begin{figure}[H]
\includegraphics[page=1]{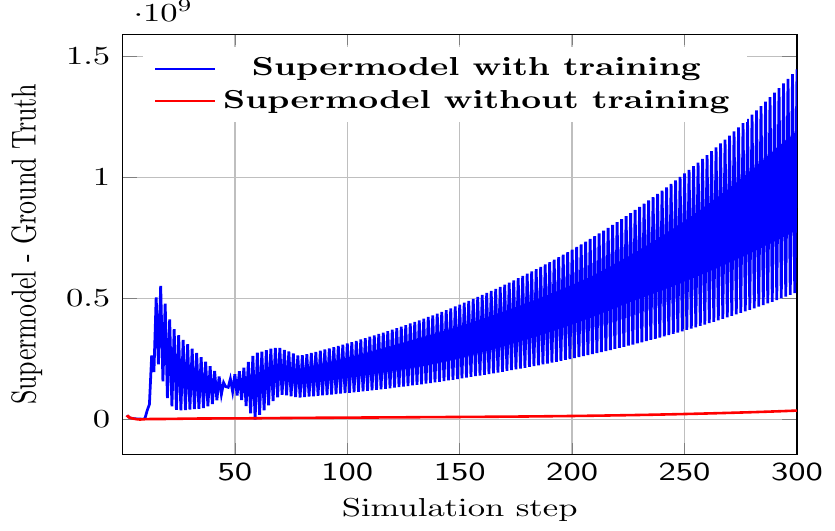}
  \caption{Difference between supermodel with respect to the GT, for the supermodel before and after the training phase.}
\label{fig:sim1g}
\end{figure}

Finally, in Figure~\ref{fig:sim1g} we present the difference between supermodel with respect to the GT, for the supermodel before and after the training phase.

All presented figures confirms that in given configuration (i.e. with $K=2.0$) supermodeling does not work and, as mentioned, the reason is too high value of the $K$ coefficient what results in numerical oscillations.

\subsection{Second experiment}
\label{sec:exp2}

\begin{figure}[h]
\includegraphics[page=1]{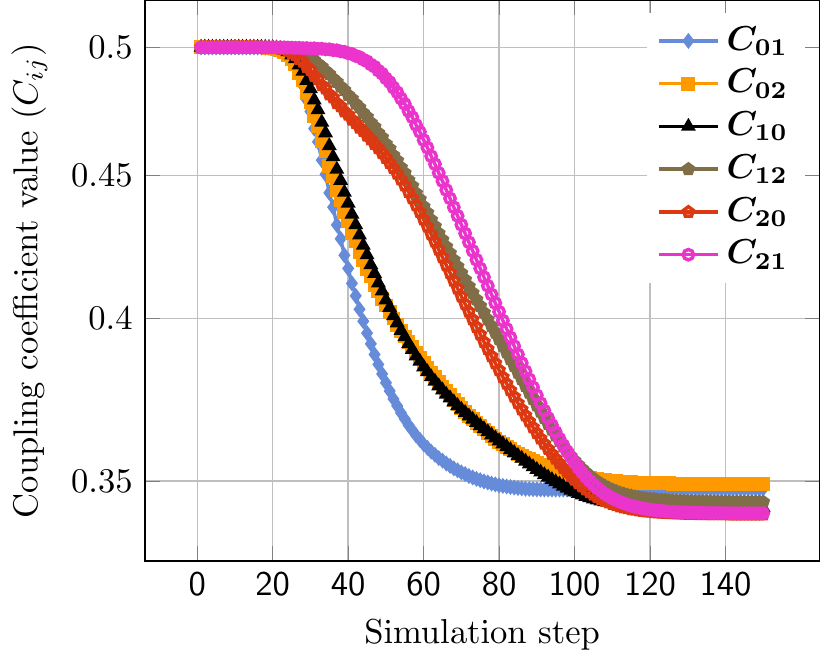}
\caption{Convergence of coupling coefficients $C_{ij}$ for ${K=0.9}$.}
\label{fig:sim2}
\end{figure}

\begin{figure}[h]
\includegraphics[page=1]{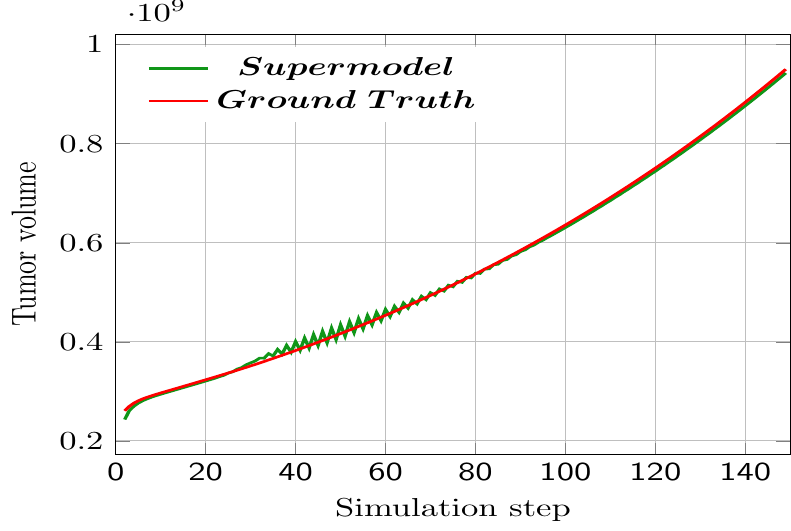}
  \caption{Convergence of tumor volumes for supermodel with respect to GT.}
\label{fig:sim2b}
\end{figure}

\begin{figure}[h]
\includegraphics[page=1]{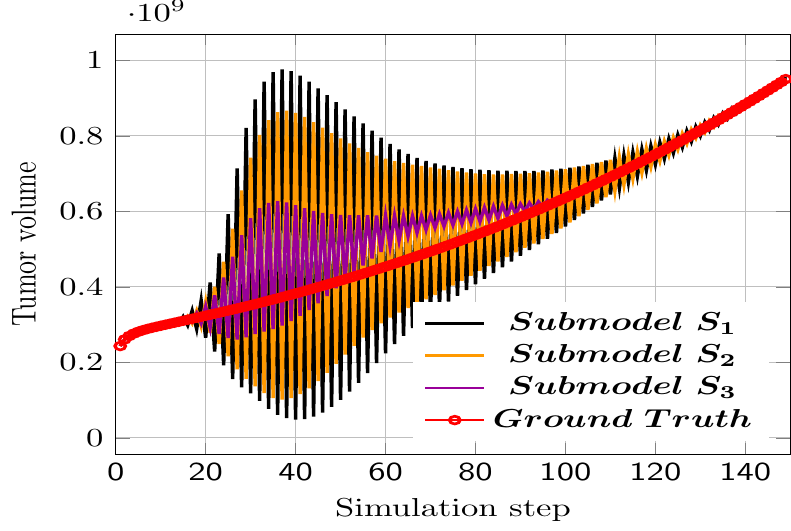}
  \caption{Convergence of tumor volumes for sub-models with respect to GT.}
\label{fig:sim2c}
\end{figure}

\begin{figure}[h]
\includegraphics[page=1]{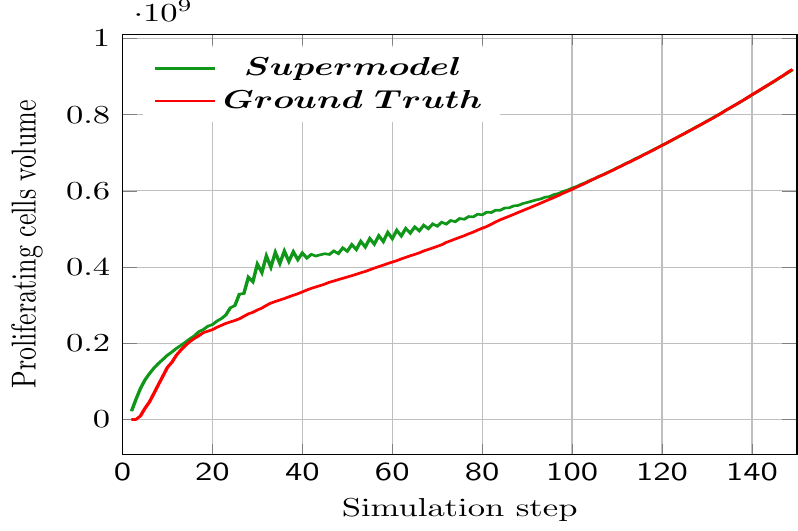}
  \caption{Convergence of the volumes of tumor proliferating cells for supermodel with respect to the GT.}
\label{fig:sim2d}
\end{figure}

\begin{figure}[h]
\includegraphics[page=1]{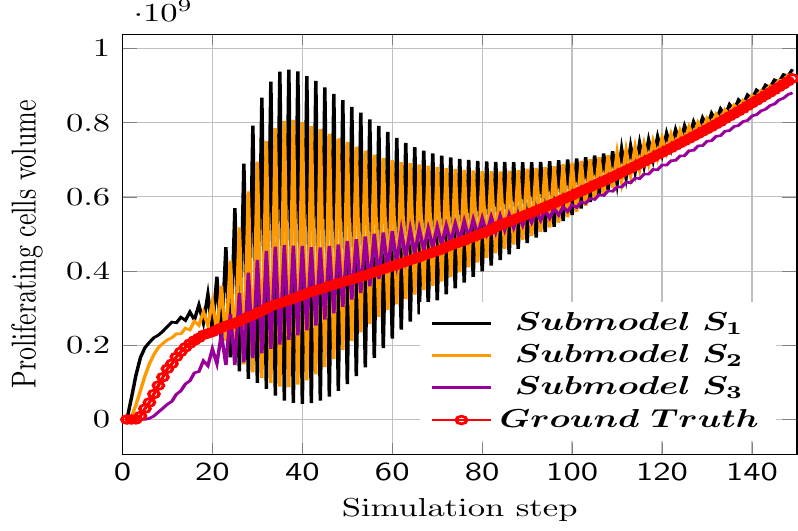}
  \caption{Convergence of the volumes of tumor proliferating cells for sub-models with respect to the GT.}
\label{fig:sim2e}
\end{figure}

\begin{figure}[h]
\includegraphics[page=1]{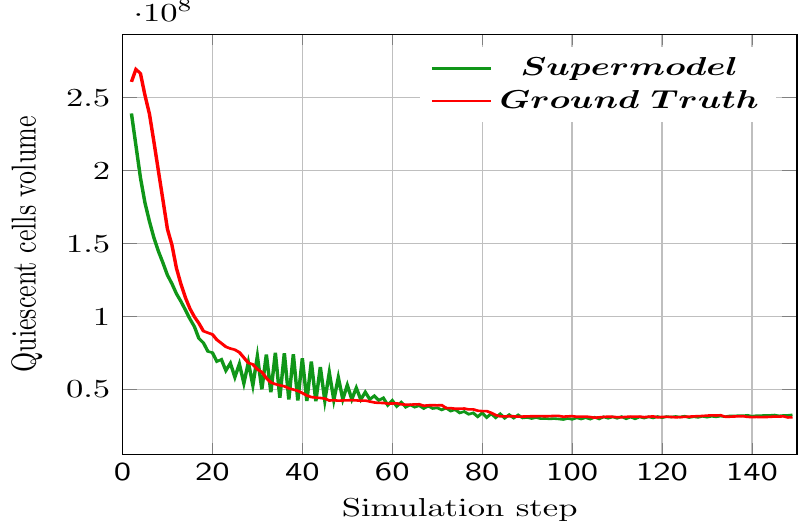}
  \caption{Convergence of the volumes of tumor quiescent cells for supermodel with respect to the GT.}
\label{fig:sim2f}
\end{figure}

\begin{figure}[h]
\includegraphics[page=1]{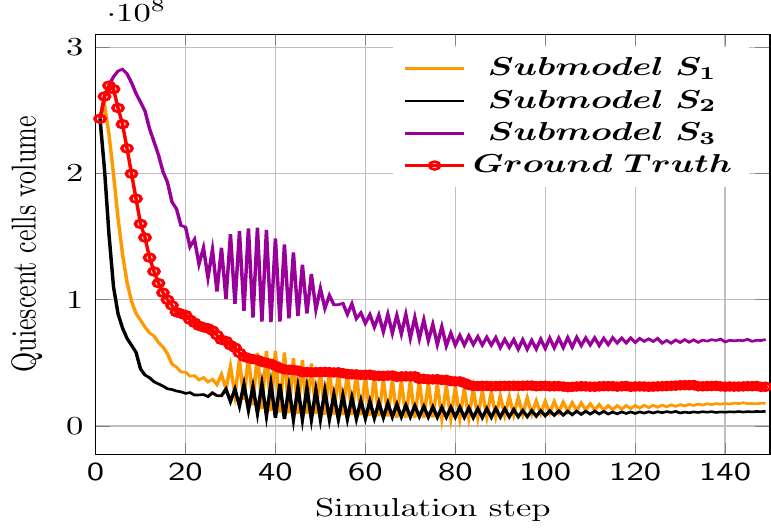}
  \caption{Convergence of the volumes of tumor quiescent cells for sub-models  with respect to the GT.}
\label{fig:sim2g}
\end{figure}

Next, we build three sub-models in the same way as before, but this time we select $K=0.9$.
Figure \ref{fig:sim2} demonstrates that the coupling coefficients $C_{ij}$ converge indeed to the reasonable value.
In Figures \ref{fig:sim2b} and \ref{fig:sim2c}, we present the convergence of the supermodel and its sub-models to the GT. We measure the convergence in terms of the total tumor volume. We present the volumes for particular sub-models (Figure~\ref{fig:sim2c}), for the average of the sub-models (Figure~\ref{fig:sim2b}) and for the GT.
The total tumor volume presented in Figure \ref{fig:sim2b} consists of the volume of proliferating and quiescent cells. We present the volumes of proliferating cells for particular sub-models (\ref{fig:sim2e}), as well as for the average of the sub-models (\ref{fig:sim2d}) and the GT (''reality'').
Next, in Figures \ref{fig:sim2f} and \ref{fig:sim2g} we present the convergence of the supermodel to the GT, this time measured in terms of the quiescent tumor cells.

We have three sub-models with the "nudging" coefficient $K=0.9$, and the coupling coefficients are $C_{ij}\approx 0.35, i,j=1,2.3$, and the time step $dt=0.1$. Thus, our constants fulfill the assumptions of the theorem, and 
\begin{eqnarray}
\alpha \leq
 \left(\inred{\left((1-K)+dt*\max_i const({\cal S},{\cal P}_i)\right)}+ 2max\{ |C^k_{ij}|,|E^k_{ij}|\}\right) = \nonumber \\
\left(\inred{0.1+0.1*\max_i const({\cal S},{\cal P}_i)}+ 2*0.35\right) =\nonumber \\0.8+0.1*\max_i const({\cal S},{\cal P}_i).
\end{eqnarray}
We do not know the exact values of the Lipschitz constants $const({\cal S},{\cal P}_i)$. They depend on the parametrizations of the PDEs. Since in our current model we do not include the treatment, the tumor volume increases with the simulation, and thus these constants are rather slightly larger than 1, but they are multiplied by the time step size $dt=0.1<1.0$.

\subsection{Disturbance of insensitive parameters}
\label{sec:insensitive}

Another interesting issue is the behavior of the supermodel when in the initial stage, the values of not only the most sensitive but also all the others parameters are disturbed.
We repeat the experiments discussed in section \ref{sec:exp2}. This time, additonally, we disturb all the other model parameters.
The disturbance of parameters was carried out in three variants - let's call them Variant I, Variant II and Variant III. In contrast, the disturbance parameter scheme presented in section \ref{sec:exp2} will be called Variant 0.
In Variant I, all other parameters were disturbed "around" the reference values with a deviation of $\pm10\%$ from these reference values. In Variant II, all other parameters were disturbed "around" the reference values with a deviation of $\pm30\%$ from the reference values. Finally, in Variant III, various schemes for distributing the parameters were used for individual sub-models. For the first sub-model, all its parameters were scattered around the reference values with a deviation of $\pm10\%$. For the second sub-model all its parameters were scattered "below" the reference values in the range of: $[ref - 20\%ref , ref-10\%ref] $. For the third sub-model all its parameters were scattered "above" reference values in the range of $[ref + 10 \% ref, ref + 20\% ref]$.

\begin{figure}[]
    \begin{tikzpicture}[yscale=.75]
     \begin{axis}[
        grid=major,
        yminorticks=false,
        xminorticks=false,
        ylabel style={font=\rmfamily},
		xlabel style={font=\rmfamily},
		xticklabel style={font=\sansmath\rmfamily},
		xlabel={Simulation step},
		ylabel={Tumor Volume Difference [GT - Supermodel]},
		legend style={draw=none,legend columns=-1},
        legend style={at={(0.5,-.3)},anchor=north},
        xmin=2,
        xmax=149
	 ]
		\addplot[restrict x to domain=2:299, line width=1pt,color=giallo]
	table[x=x,y=e1b] {ressec4.txt};

	\addplot[restrict x to domain=2:299, line width=1pt,color=black]
	table[x=x,y=e1a] {ressec4.txt};

		\addplot[restrict x to domain=2:299, line width=1pt,color=verde]
	table[x=x,y=e1d] {ressec4.txt};

	\addplot[restrict x to domain=2:299, line width=1pt,color=viola]
	table[x=x,y=e1c] {ressec4.txt};

     \legend{ Variant 0,Variant I,Variant III,Variant II}

    \end{axis}
        \node [anchor=north east,color=darkgray,inner sep=0,xshift=-190pt,yshift=-25pt]
      at (current page.south) {\bf{Parameters disturbance scheme:}};
    
\end{tikzpicture}
  \caption{Comparison of supermodel convergence for different initial (insensitive) parameter disturbance schemes for  ${K=0.9}$}
\label{fig:e1}
\end{figure}

As one can see in charts presented in Figure~\ref{fig:e1} for all three additional parameters disturbance schemes, the supermodel was able to converge towards the GT, but this time it took a little bit longer than for Variant 0. 
The disturbed sub-models have possibly larger $\alpha$ constant, since they depend on the Lipschitz constant, and the Lipschitz constant depends on parameters ${\cal P}_i$, the constants may be larger, so the rate of convergence will be slower, as results from Remark 3.







    

\subsection{Dependence on the number of sub-models and their parametrization}
\label{sec:numberofsubmodels}

In the following experiments, we test the convergence of the supermodeling based on different number of sub-models and different initial parameters distributions. 
We select as the reference oxygen proliferation parameter as $o^{prol}=10$. We construct different sub-models with reference to this value.

	We investigate this aspect by performing some experiments where we initially disperse sub-models parameters in three different ways, i.e.
\begin{itemize}
	\item The parameters of sub-models have been distributed evenly around the reference value.  
	\item The parameters of sub-models have been distributed  ''below'' the reference value. 	
	\item Parameters of the sub-models have been distributed  ''above'' the reference value. 
\end{itemize}

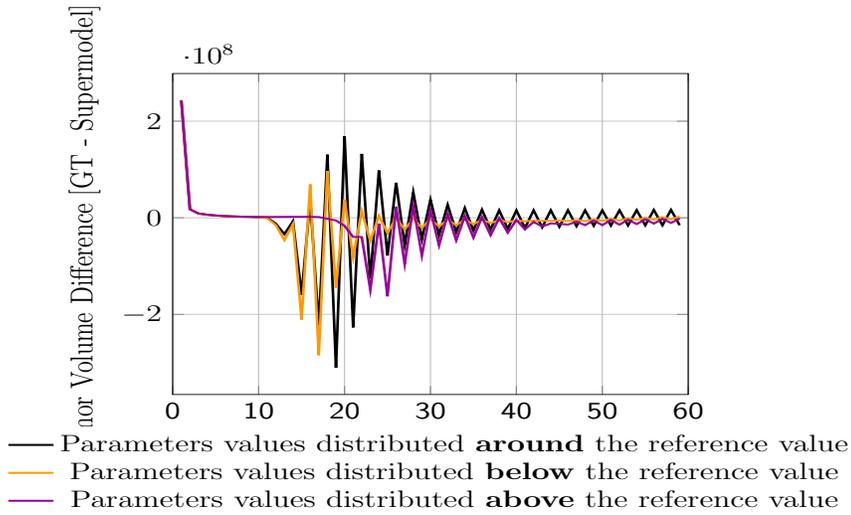
\begin{figure}[]
	\begin{tikzpicture}[yscale=.75]
		\begin{axis}[
			grid=major,
			yminorticks=false,
			xminorticks=false,
			ylabel style={font=\rmfamily},
			xlabel style={font=\rmfamily},
			xticklabel style={font=\sansmath\rmfamily},
			xlabel={Simulation step},
			ylabel={Tumor Volume Difference [GT - Supermodel]},
			legend style={draw=none},
			legend style={at={(0.5,-0.1)},anchor=north},
			xmin=0,
			xmax=60
			]
			\addplot[restrict x to domain=1:59, line width=1pt,color=black]
			table[x=x,y=N7] {diffs678.txt};
			
			\addplot[restrict x to domain=1:59, line width=1pt,color=giallo]
			table[x=x,y=B7] {diffs678.txt};
			
			\addplot[restrict x to domain=1:59, line width=1pt,color=viola]
			table[x=x,y=A7] {diffs678.txt};

			\legend{Parameters values distributed \textbf{around} the reference value, Parameters values distributed \textbf{below} the reference value, Parameters values distributed \textbf{above} the reference value}
		\end{axis}
	\end{tikzpicture}
	\caption{The difference between {GT} and supermodel predicted tumor volume for super model consisting of {7} sub-models and with different starting parameter values distributions.}
	\label{fig:diff7}
\end{figure}

\begin{figure}[]
	\begin{tikzpicture}[yscale=.75]
		\begin{axis}[
			grid=major,
			yminorticks=false,
			xminorticks=false,
			ylabel style={font=\rmfamily},
			xlabel style={font=\rmfamily},
			xticklabel style={font=\sansmath\rmfamily},
			xlabel={Simulation step},
			ylabel={Tumor Volume Difference [GT - Supermodel]},
			legend style={draw=none},
			legend style={at={(0.5,-0.1)},anchor=north},
			xmin=0,
			xmax=60
			]
			\addplot[restrict x to domain=1:59, line width=1pt,color=black]
			table[x=x,y=N8] {diffs678.txt};
			
			\addplot[restrict x to domain=1:59, line width=1pt,color=giallo]
			table[x=x,y=B8] {diffs678.txt};
			
			\addplot[restrict x to domain=1:59, line width=1pt,color=viola]
			table[x=x,y=A8] {diffs678.txt};

			\legend{Parameters values distributed \textbf{around} the reference value, Parameters values distributed \textbf{below} the reference value, Parameters values distributed \textbf{above} the reference value}
		\end{axis}
	\end{tikzpicture}
	\caption{The difference between {GT} and supermodel predicted tumor volume for super model consisting of {8} sub-models and with different starting parameter values distributions.}
	\label{fig:diff8}
\end{figure}
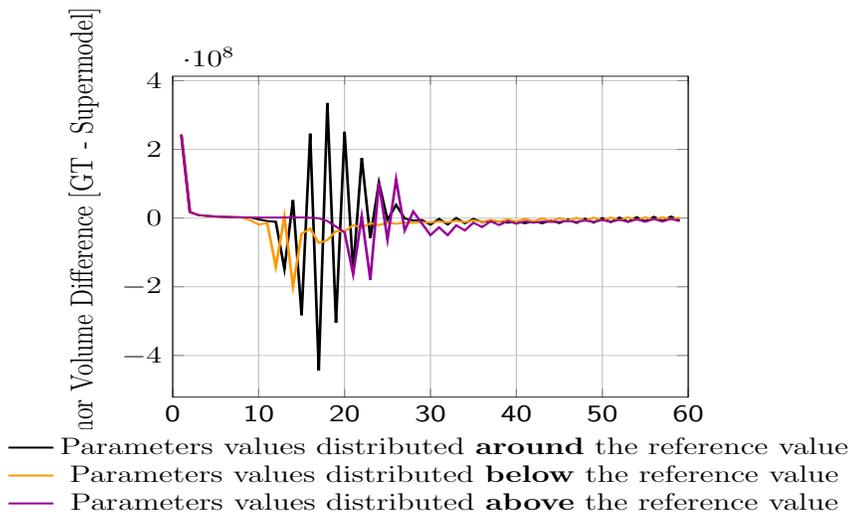

In Figures~\ref{fig:diff6}-\ref{fig:diff8} we present some exemplary results. We demonstrate the results for three supermodels consisting of six, seven, and eight sub-models, respectively.
	For experiments with six sub-models, we took the following configurations:
\begin{itemize}
	\item In three of sub-models the $o^{prol}$ parameter value has been set below the reference value. The other three of the sub-models have this parameter set above the reference value.
	
	\item In all sub-models the $o^{prol}$ parameter has the value below the reference value.
	
	\item In all sub-models the $o^{prol}$ parameter  has the value above its reference value. 
\end{itemize}

\begin{figure}[]
	\begin{tikzpicture}[yscale=.75]
		\begin{axis}[
			grid=major,
			yminorticks=false,
			xminorticks=false,
			ylabel style={font=\rmfamily},
			xlabel style={font=\rmfamily},
			xticklabel style={font=\sansmath\rmfamily},
			xlabel={Simulation step},
			ylabel={Tumor Volume Difference [GT - Supermodel]},
			legend style={draw=none},
			legend style={at={(0.5,-0.1)},anchor=north},
			xmin=0,
			xmax=60
			]
			\addplot[restrict x to domain=1:59, line width=1pt,color=black]
			table[x=x,y=N6] {diffs678.txt};
			
			\addplot[restrict x to domain=1:59, line width=1pt,color=giallo]
			table[x=x,y=B6] {diffs678.txt};
			
			\addplot[restrict x to domain=1:59, line width=1pt,color=viola]
			table[x=x,y=A6] {diffs678.txt};

			\legend{Parameters values distributed \textbf{around} the reference value, Parameters values distributed \textbf{below} the reference value,Parameters values distributed \textbf{above} the reference value}
		\end{axis}
	\end{tikzpicture}
	\caption{The difference between {GT} and supermodel predicted tumor volume for super model consisting of {6} sub-models and with different starting parameter values distributions.}
	\label{fig:diff6}
\end{figure}

{Similarly, for experiments with seven sub-models, we took the following configurations:} 
\begin{itemize}
	\item 
In four of sub-models the $o^{prol}$ parameter value has been set below the reference value. In the other three sub-models this parameter has values above the reference value.
	
	\item In all sub-models the $o^{prol}$ parameter has the value below the reference value .
	
	\item In all sub-models the $o^{prol}$ parameter  has the value above its reference value.
\end{itemize}

{Finally, for experiments with eight sub-models, we took the following configurations: }
\begin{itemize}
	\item 
In four of sub-models the $o^{prol}$ parameter value has been set below the reference value. In the other sub-models this parameter has been set above the reference value.
	
	\item In all sub-models the $o^{prol}$ parameter has the value below the reference value.
	
	\item In all sub-models the $o^{prol}$ parameter has the value above its reference value.
\end{itemize}

	We summarize results in Figures~\ref{fig:diff6}-\ref{fig:diff8}. Larger number of sub-models and distribution of parameters, different oscillations, and convergence. Different values for the $o^{prol}$ results in different Lipschitz constants in $\alpha$ and by varying the parameter we can effectively change the Lipschitz constant. Increasing number of sub-models makes it possible that some of them are worse, having larger Lipschitz constant value, which may result in different convergence.

\section{Conclusions}
In this paper, we concentrate on studying the supermodel's convergence aspects, i.e., dynamically synchronized ensemble of sub-models. To this end, first, we have theoretically proved its conditional convergence and collect the most important convergence principles. 

We have proved that the supermodel convergence depends on the quality of the sub-models initialization (the parameters that instantiate the PDEs and the resulting Lipschitz continuity constant), their distribution in the phase space, and their diversity (which is related to the distance to the GT, and again the Lipshitz continuity constants). To demonstrate the supermodel convergence experimentally, we show how it works for a non-trivial 3D model of tumor growth described by a set of PDEs. 

From the theoretical estimates, we concluded that 
(1) the "nudging'' of the supermodel to the GT has to be well-balannced to make the system stable. 
(2) the PDE operator modeling the evolution needs to be bounded, with the convergence of the training controlled by the time-step size, and 
(3) all the sub-models have to be relatively close but not too close to GT for their entire trajectory.
Based on a complex tumor evolution model, we show that just by training several meta-parameters of the supermodel ($C_{ij}$ and $K$), we can fit the GT data in a low computational time. This training takes around one hour of computations on one node of Prometheus Linux cluster from CYFRONET supercomputing center \cite{Wiatr}. The application of classical data assimilation for such a complex model with more than 20 parameters is extremely computationally demanding due to the exponential inflation of parameter space with their number.

In our future work, we plan to extend the tumor model into the equations modeling the treatment therapy and run the supermodeling algorithm based on medical data. Then, such the cancer supermodel would be used as the computational engine in prognostic oncology in planning anti-cancer therapy.
\section*{Acknowledgments}
The work has been suported by Polish National Science Centre, Poland grant no. 2016/ 21/B/ST6/01539,  and in part by PL-Grid Infrastructure. The visit of Maciej Paszy\'{n}ski at Oden Institute, The University of Texas at Austin, USA has been supported by J. T. Oden Research Faculty Fellowship. Maciej Paszy\'nski would like to thank Prof. Jean-Luc Guermond from Texas A\& M for full of insights discussion on the convergence of the supermodeling.

\end{document}